# Experimental Demonstration of Degenerate Band Edge in Metallic Periodically-Loaded Circular Waveguide

Mohamed A. K. Othman, Xuyuan Pan, Georgios Atmatzakis, Christos Christodoulou, and Filippo Capolino

*Abstract*—We experimentally demonstrate for the first time the degenerate band edge (DBE) condition, namely the degeneracy of four Bloch modes, in loaded circular metallic waveguides. The four modes forming the DBE represent a degeneracy of fourth order occurring in a periodic structure where four Bloch modes, two propagating and two evanescent, coalesce. It leads to a very flat wavenumber-frequency dispersion relation, and the finite-length structure's quality factor scales as $N^5$ where $N$ is the number of unit cells. The proposed waveguide in which DBE is observed here is designed by periodically loading a circular waveguide with misaligned elliptical metallic rings, supported by a low-index dielectric. We validate the existence of the DBE in such structure using measurements and we report good agreement between full-wave simulation and the measured response of the waveguide near the DBE frequency; taking into account metallic losses. We correlate our finding to theoretical and simulation results utilizing various techniques including dispersion synthesis, as well as observing how quality factor and group delay scale as the structure length increases. Moreover, the reported geometry is only an example of metallic waveguide with DBE: DBE and its characteristics can also be designed in many other kinds of waveguides and various applications can be contemplated as high microwave generation in amplifiers and oscillators based on an electron beam interaction or solid state devices, pulse compressors and microwave sensors.

## I. INTRODUCTION

DEGENERACY conditions refer to points in the dispersion diagram of guided modes in a periodic structure at which Bloch eigenmodes coalesce. Those points of degeneracy can be found in lossless waveguides at the cutoff frequencies, or at the band edge of periodic structures [1]–[6]. A regular band edge (RBE) is a standard degeneracy condition that occurs in all periodic guiding structure at the edge of the Brillouin zone where forward and backward harmonics of a propagating Bloch mode coalesce, between a stop and a pass band. The dispersion relation of the Bloch wavenumber $k$ and angular frequency $\omega$ of the modes contributing to the RBE, for instance, is equivalent

This work was supported by AFOSR MURI Grant No. FA9550-12-1-0489 administered through the University of New Mexico. M. O. and F. C. also acknowledge support from AFOSR Grant No. FA9550-15-1-0280.

M. O. and F. C. are with the Department of Electrical Engineering and Computer Science, University of California at Irvine, Irvine, CA 92697 USA (e-mail: mothman@uci.edu, f.capolino@uci.edu).

X. P., G. A. and C. C. are with the Department of Electrical and Computer Engineering, University of New Mexico, Albuquerque, NM 87131 USA (e-mail: christos@unm.edu).

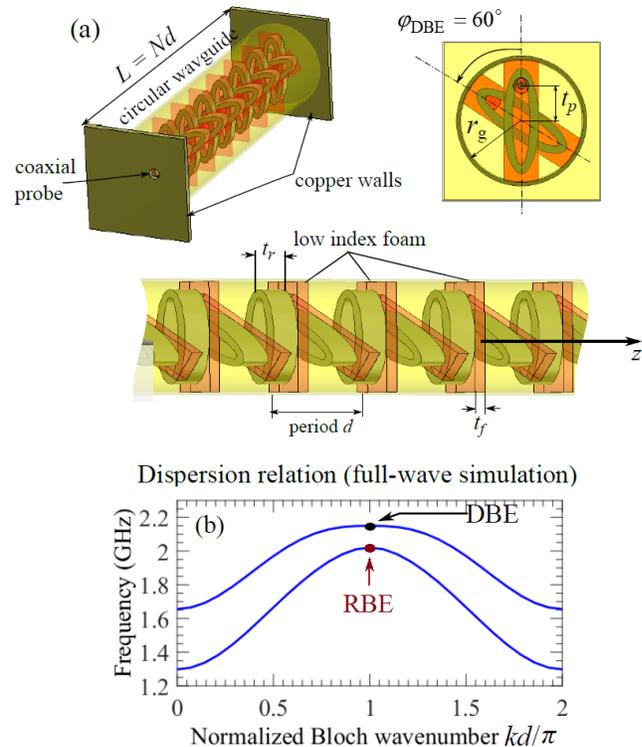

Fig. 1. (a) Geometry of a periodically-loaded circular waveguide under study that exhibits a DBE. The circular waveguide is loaded with elliptical rings with a designed misalignment angle of $\varphi_{DBE}$=60 degrees. The rings are supported by a low-index dielectric that is not affecting mode propagation. A coaxial probe is inserted on a metallic wall at a distance $t_p$ from the waveguide axis. (b) Corresponding dispersion diagram obtained from full-wave simulations: the DBE is obtained at ~2.16 GHz.

to $(\omega_g - \omega) \propto (k - k_d)^2$ where $\omega_g$ and $k_g = \pi/d$ are the RBE angular frequency and wavenumber, respectively, with $d$ being the period of the periodic waveguide. The band edge condition is normally associated with slow-wave properties in the sense that group velocity exhibits substantial reduction near the band edge [7]–[10]. Resonators made of waveguides that support modes with extremely low group velocity possess very high quality ($Q$) factor [3], [9], [11]–[13] and therefore are very suitable to applications including filters, pulse forming networks, in high power microwave generation [14] and in RF accelerators [15].

The purpose of this paper is to investigate and provide the first experimental verification of a fourth order degeneracy among four Bloch modes occurring at the band edge in a metallic, periodically-loaded hollow waveguide. This condition is called *degenerate band edge* (DBE) condition and has been studied previously in special anisotropic dielectric stacks [3]. Moreover, DBE has been also investigated theoretically in the literature [3], [5], [9], [12], [14]–[16]. Indeed, the DBE has been shown to provide unprecedented enhancement in the *Q* factor and field enhancement over RBE structures as demonstrated theoretically in [10], [19]. Furthermore, applications featuring slow-light properties associated with the DBE has been proposed such as small/directive antennas [12], [20]–[23], low-threshold switching [10], [13], pulse compression [24], and filters [11]. The theory of four mode degeneracy was furthermore utilized for the first time to show giant gain enhancement [10] in slow-wave structures with DBE, for high power amplifiers [17] and oscillators [18], [25], when interacting with an electron beam. Since DBE involves four degenerate modes, it requires sophisticated theoretical framework to reveal its unique properties; such as coupled transmission line theory developed in [3], [6], with a non-diagonalizable system matrix. DBE can be designed in various technologies such as coupled microstrip lines [5], [26], stacks of laminates layers of printed strips [21], and artificial anisotropic dielectrics [23], [22]. However, in connection with experimental studies, the giant resonance (as was referred to in [3], [27]), and giant *Q* factor scaling associated with the DBE was shown only in optical waveguides [28], [29]. In fact, DBE dispersion, and *Q* factor scaling as well as the effect of losses have never been shown experimentally at microwaves.

In this paper we (i) demonstrate for the first time the band edge degeneracy as well as giant scaling of the *Q* factor in metallic waveguides at microwave frequencies; stimulated by its intriguing physics and various applications, including pulse compression devices [24] and high power microwave generation [6], [17], [18]. (ii) Illustrate how DBE and its properties can be implemented methodologically in metallic waveguide loaded with elliptical rings; for which the design and analysis were reported in [18], [30], (iii) Show from both full-wave simulations and measurements that the structure indeed supports a band edge degeneracy even in the presence of metal losses; though losses may obscure the observation of such degeneracies, especially in long guiding structures. We stress that, although the reported experimental study pertains to a specific waveguide geometry, DBE is a general property that can be deliberately attained in various other topologies of all-metallic waveguides that support at least two modes in each direction.

## II. LOADED WAVEGUIDE WITH DBE: DESIGN AND SIMULATIONS

In Fig. 1 we illustrate the geometry of the periodic waveguide under study that exhibits a DBE based on the theoretical demonstration in [30]. Here the period is *d*=35 mm and the unit cell comprises a circular waveguide operating below cutoff, with radius of $r_g$ = 35 mm loaded with two elliptical metallic rings, each with a thickness of $t_r$=9 mm in the longitudinal direction and aspect ratio of 2:1 (ratio between major and minor radii of the elliptical ring). The ring inner (outer) major radius is 25 mm (30 mm) and the ring's separation is $t_f$ =4 mm The ring's major axes are misaligned by $\varphi_{DBE}$ = 60 degrees; which is one of the crucial design parameters [18], [30]. The rings are floating (not in electric contact with the waveguide wall) and they are supported by dielectric rectangular fixtures as seen in Fig. 1(a), whose relative dielectric constant is 1.05. Such dielectric supports have no tangible effect on the waveguide propagation characteristics since we use a very low index foam as seen next. The DBE frequency is 2.16 GHz. The design rules of such structure, and a thorough analysis using full-wave simulations and coupled transmission line model are discussed in [18], [30]. Moreover, the role of the misalignment angle is rather very critical for the DBE to manifest [18], [30]. The elliptical rings provide means for the anisotropic behavior (in the *x-y* plane) and therefore are able to support two polarizations that are periodically coupled while propagating along the *z*-direction. An analogous concept was discussed in [31] but in a different geometry exhibiting the split band edge (SBE) [3], [32]. Since each polarization implies two modes, one in each direction due to the ±*k* symmetry in reciprocal structures, the coupling allows to develop a DBE, i.e., a degeneracy between four modes at the band edge. We use a full-wave method, i.e., the finite element method (FEM) eigenmode solver implemented in CST Microwave Studio, to calculate the dispersion of our proposed periodic structure, by simulating its unit cell using periodic boundary conditions, with a phase shift equal to $kd$ (in radian). The derivatives $\partial^n \omega / \partial k^n$ at the band edge, i.e., when $k = \pi/d$, for $n = 1, 2, 3$ and 4 for different $\varphi$ values are then calculated numerically. As such, we identify an angle $\varphi = \varphi_{DBE}$ at which the first three derivatives vanish while the fourth derivative is non-zero $\partial^4 \omega / \partial k^4 \neq 0$. This designates the condition under which DBE dispersion relation $(\omega_d - \omega) \approx h(k - k_d)^4$ is fulfilled [3], [30], where $h$ is a constant that depends on the geometry and will be given in Section III.B for the waveguide under consideration. Note that for $\varphi = 0°$ or $90°$, a DBE condition can never be met, for any other parameter choices, due to insufficient polarization mixing. The particular misalignment angle $\varphi$ in the range $0° < \varphi < 90°$ that enables the DBE depends on the choice of the geometrical parameters. For the implemented waveguide, a DBE condition is found here when $\varphi = \varphi_{DBE} \cong 60°$. Incidentally, small variations around such value of $\varphi = \varphi_{DBE} \cong 60°$ is convenient for tuning purposes [18], [31], as shown later in Section III.C.

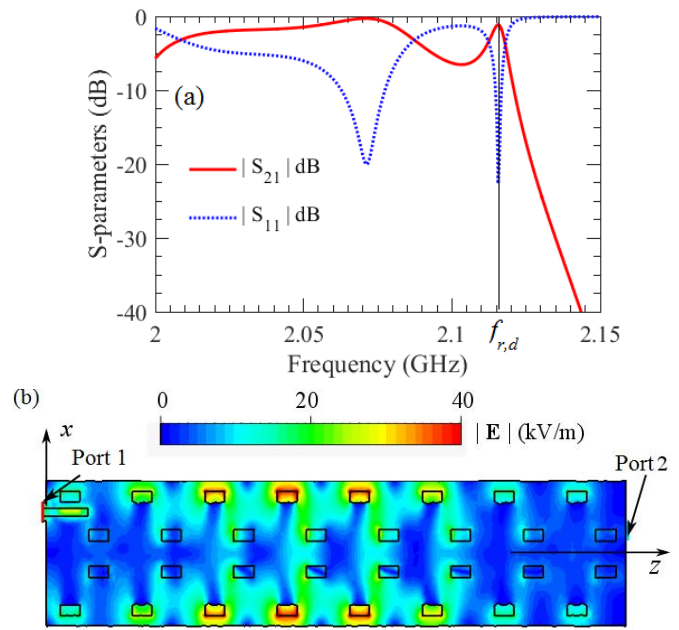

Fig. 2. (a) S-parameters of the DBE waveguide in Fig. 1 with *N*=8 unit cells, terminated on ports made by coaxial probes. (b) Magnitude of the total electric field in the *x-z* plane at the DBE resonance occurring at $f_{r,d}$ = 2.129 GHz (that is the last resonance peak occurring at a frequency $f_{r,d}$ lower than the DBE in (a)), when the structure is excited with an incident power of 1 W. Results are obtained via full-wave simulations.

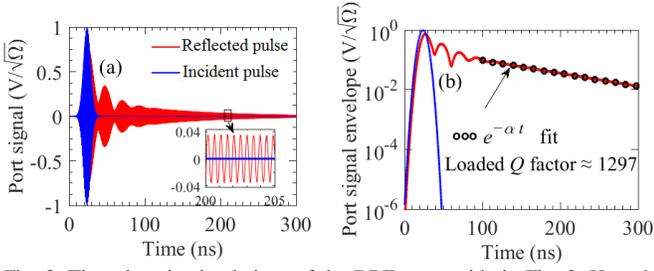

Fig. 3. Time-domain simulations of the DBE waveguide in Fig. 2. Here the waveguide is made of $N = 8$ unit cells and terminated with a "short" on Port 2. (a) Incident Gaussian pulse on Port 1 and the corresponding reflected pulse at the same port. (b) Envelope of the incident and reflected pulses, and a fitting of exp(-$\alpha t$) to the pulse tail is shown, which is used to estimate the loaded $Q$ for the cavity. The loaded $Q$ factor of the cavity for $N=8$ unit cells with "short" termination on Port 2 is ~ 1297 which agrees with the experimental results in Fig. 4 and Fig. 5.

In Fig. 1 the dispersion diagram of the two lowest order modes is depicted (both forward and backward harmonics since the Brillouin zone extends from $k = 0$ to $k = 2k_d = 2\pi/d$). The dispersion diagram exhibits a DBE at $f_d = \omega_d/(2\pi) \approx 2.16$ GHz that can be well fitted by $(\omega_d - \omega) \approx h(k - k_d)^4$. Moreover, the lower order mode (forward and backward harmonic) exhibit a regular band edge (RBE) at ~1.95 GHz.

A finite length resonator based on cascading a number $N$ of unit cells described above is constructed by placing two copper walls at both ends of the waveguide with a through hole of radius 4.5 mm for the probe that excites the DBE mode as discussed in [18], see Fig. 1. Coaxial probes on each side of the resonator (i.e, Ports 1 and 2) are located at a distance $r_p$=20 mm from the waveguide axis as shown in Fig. 1, with the inner conductor of radius of 2 mm, and with the through hole radius of 4.5 mm. The coaxial probe has a length of 20 mm inside the waveguide. The DBE waveguide with termination is made of copper whose conductivity is taken as $5.8 \times 10^7$ S/m. Since the group velocity of the DBE modes is vanishing, a giant Fabry-Perot (FP) resonance mode is formed near the DBE frequency, at which the Q-factor and group delay are very high [10]. The simulated S-parameters of the DBE cavity excited by the coaxial probes are numerically obtained by FEM implemented in CST Microwave Studio and are depicted in Fig. 2 for a structure with $N = 8$ unit cells. Since a bandgap occurs at higher frequencies [i.e., for $f > f_d = \omega_d/(2\pi)$ ] it is clear that for frequencies higher than the last transmission resonance at ~2.129 GHz the transmitted field drops to very low values, as evident from Fig. 2(a) where the transmission coefficient ($S_{21}$) rapidly decreases to −40 dB. The transmission peak of interest occurs at a frequency slightly lower than the DBE frequency $f_d$, and is designated as the DBE resonance of the finite-length resonator denoted by $f_{r,d}$ [3], [10], [12]. The DBE cavity is well-matched at the resonance frequencies ($|S_{11}| < -20$ dB) whereas the magnitude of transmission is rather high ($|S_{21}| > -3$ dB) considering that losses are taken into account in the full-wave simulations and they strongly affect the DBE performance. The characteristics of such DBE resonance in analogous periodically-loaded structures and how its frequency and linewidth scale with the structure length is shown theoretically in [3], [10], [19], [20], [28] and will be revisited in Section III. We also report the magnitude of the total electric field in the x-z plane cut along the cavity when excited from one port on the left in Fig. 2 using the coaxial probe configuration described above, at the DBE resonance frequency $f_{r,d} = 2.129 \text{ GHz}$. A strong field enhancement in the DBE cavity is observed, resulting in up to $2 \times 10^4$ V/m of field strength near the center of the cavity at the DBE resonance (for 1 W of average incident power, i.e., source power, from Port 1 at $f_{r,d} = 2.129 \text{ GHz}$ for which $|S_{11}| < -20$ dB). High field enhancement is a clear signature of the frozen mode regime (as was referred to in [27]) when operating at the resonating frequency, even in the presence of metal losses [18].

To provide an estimation of the loaded $Q$ factor of the DBE resonance, we perform a time-domain simulation using the time-domain full-wave solver implemented in CST Microwave Studio. The resonant cavity is excited by a coaxial probe at Port 1, as discussed previously, and terminated by a "short" on Port 2. The input signal is a Gaussian pulse whose center frequency is 2.1 GHz and the full-width half-maximum (FWHM) is 0.2 GHz and is shown in Fig. 3. The resulting reflected signal at Port 1 displays the feature of the high $Q$-factor DBE resonance in which the fields take a large amount of time to decay, and the tail of the output pulse comprises of only a single frequency, namely the DBE resonance $f_{r,d}$. Moreover, we plot in Fig. 3(b) the envelope of the output pulse and perform a fitting of the pulse tail to $\exp(-\alpha t)$ where $\alpha$ is the damping constant of the cavity and is related to the $Q$ factor by $\alpha = \pi f_{r,d}/Q$ as typical in resonant cavities (see p. 388 in [33], and [34]). Accordingly, we find that, for the case of $N = 8$ unit cells, $\alpha \cong 5.15 \, (\mu s)^{-1}$ which leads to a loaded Quality factor of $Q \approx 1297$. We will show in Section III a good agreement between this estimated $Q$ factor and the measured one. The focus of the rest of this paper is to show that the DBE is experimentally demonstrated for the waveguide geometry in Fig. 1 and compare simulation predictions and measurements.

## III. LOADED WAVEGUIDE WITH DBE: FABRICATION AND TEST

### A. Waveguide Fabrication

The rings are fabricated from copper (copper grade 110 which in theory has a conductivity of ~$5.8\times10^7$ S/m) using a waterjet cutter method. Each ring is attached to a rectangular shaped, very low-index closed-cell foam using non-conductive epoxy adhesive. The foam supports are 70×25 mm$^2$ in cross sectional area, 4 mm in thickness, and have an approximate dielectric constant of ~1.05 at 2 GHz and hence should not affect propagation characteristics. The rings with foam fixtures are individually inserted into the waveguide, and then secured to the waveguide walls using an adhesive. Notice the importance of aligning rings along the wave propagation direction (*z*-direction); which is achieved manually using measuring rods. The waveguide is terminated on both sides with copper plates, electrically connected to the metallic waveguide. The copper plates have a through hole for the coaxial probes as shown in Fig. 4, and their dimensions are described in Section II. The probe is made of a 50 Ohm coaxial connector seen in Fig. 4(a) that is affixed to the copper plate. Waveguides of different lengths were fabricated and here we report measurements pertaining to two different setups: Case 1) reflection measurements, with only one port excitation, while the other end is either (i) "shorted" meaning it is terminated by a copper plate, or (ii) "open" where the waveguide is open-ended from one side, and (iii) "50 Ohm" meaning that the waveguide is shorted with a copper wall but with a 50 Ohm coaxial probe terminated by a broadband load. Case 2) Transmission measurements when two port parameters are measured from both ends of the waveguide, i.e., reflection and transmission measurements utilizing the probes described above. The measurements were done in frequency domain using Keysight N5247A PNA-X as well as Rhode & Schwarz ZVA 67 vector network analyzers.

TABLE I: RESONANCE FREQUENCIES $f_{r,d}$ OBTAINED FROM FULL-WAVE SIMULATIONS AND MEASUREMENTS FOR A OF PERIODIC WAVEGUIDE MADE OF $N = 4$ UNIT CELLS

| | Resonance frequencies (GHz) | | | | |
|---|---|---|---|---|---|
| Full-wave simulations | 1.638 | 1.8971 | 1.98 | 2.075 | 2.147 |
| Measurements | 1.649 | 1.913 | 1.985 | 2.09 | 2.15 |

### B. Reflection measurements

We report first the one port S-parameter, namely $S_{11}$ of the waveguide excited from one end using the coaxial probe with characteristic impedance of 50 Ohm while the other end has port 2 terminated with a "short" circuit. In Fig. 4(b) we show the magnitude of $|S_{11}|$ in dB in the frequency range from 1.6

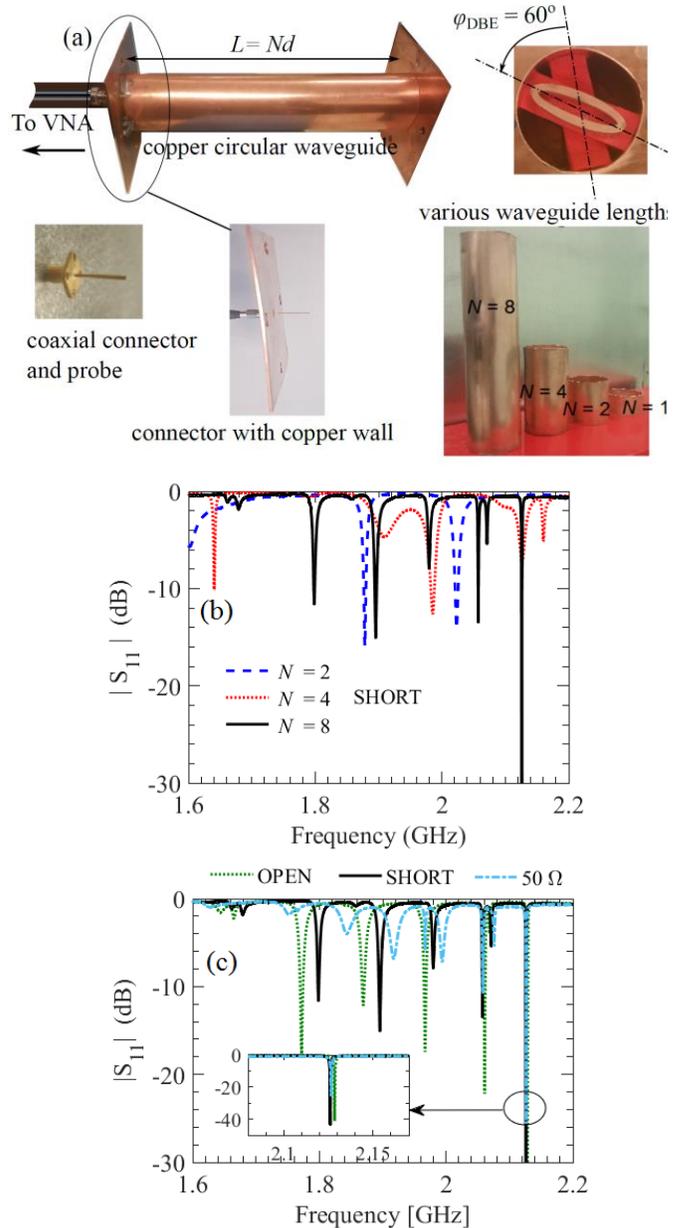

Fig. 4. (a) Fabricated metallic waveguide whose modes exhibit a DBE; showing the metallic circular waveguide, coaxial probes, the elliptic rings supported by low-index foam, and various lengths considered. (b) Measurement of the magnitude of $S_{11}$ for three different lengths of the waveguide, excited by a coaxial probe from the left (Port 1). The other right end of the waveguide (port 2 is terminated by a metallic "short" as seen in (a). (c) Measurement of the magnitude of the $S_{11}$ for three different coaxial probe terminations on port 2, namely "short", open and a 50 Ohm load, all for a length of $N = 8$ unit cells.

GHz to 2.16 GHz which is the range of frequency of interest relevant to the observation of the DBE, as seen from the dispersion in Fig. 1, for three different resonator lengths. Various resonances are observed in Fig. 4(b) as dips in the magnitude of $S_{11}$. A band gap exists beyond ~2.16 GHz, in which $|S_{11}|$ level is close to 0 dB for different lengths of the waveguide, $N= 2, 4,$ and 8. The *band edge resonance at* $f_{r,d}$, namely the dip in the $|S_{11}|$ curve that is closest to 2.16 GHz is the most pronounced feature of the waveguide. The value of the resonance frequency $f_{r,d}$ depends on the resonator length and it follows the trend $2\pi f_{r,d} \approx 2\pi f_d - h[\pi/(Nd)]^4$, that also shows that it tends to the DBE frequency for long lengths [3], [18].

In Table I we show the resonance frequencies of the waveguide resonator based on both full-wave simulations and measurements which are in good agreement, for the case with $N = 4$ cells, and a short circuit at port 2. Similar agreement is observed for both $N = 2$ and $N = 8$ and not shown here for the sake of brevity. Recall that a conventional periodic structure of $N$ periods exhibits discrete resonant frequencies with phase shifts per period spaced between 0 and $\pi$ pertaining to a sole excited mode [35]. This is depicted from the resonances in Fig. 4(b) for different lengths. It is important to point out the such setup is typically used to measure the intrinsic $Q$ of a cavity under critical coupling [1]. As such, the cavity is critically coupled at the DBE resonance as seen from the level of $|S_{11}|$ in Fig. 4(b) which is lower than $-30$ dB, when port 2 is shorted, for the DBE resonance whose frequency is $f_{r,d} \simeq 2.127$ GHz when $N = 8$.

Another important aspect of the DBE resonance, is that it is not very sensitive to load variations. Indeed, the field distribution plotted Fig. 2(b) shows that the load conditions on both ends of the waveguide would not significantly perturb the resonance field inside the DBE cavity since the field strength at the cavity ends is much smaller than the field at the center. This is also experimentally observed in Fig. 4(c) where the DBE resonance $f_{r,d}$ of the finite length resonator (the closest to the DBE frequency in Fig. 1) is essentially the same for the three different loads. Note that the other resonance frequencies of the cavity away from the DBE are different for different loads (short, open and 50 Ohm terminations). Instead, the DBE resonance closest to 2.16 GHz is not prone to variations in the load conditions. (Here, the 50 Ohm termination is achieved by using a coaxial probe at the other end of the waveguide, identical to the one used to excite the structure).

The purpose of our study is not only to show a high-$Q$-factor cavity but rather to demonstrate the DBE feature that has not been experimentally shown before in all-metallic waveguides. Therefore, the most unique DBE characteristic of a periodic waveguide is its respective dispersion diagram which portrays how the phase changes with frequency over the passband of the structure. Although different approaches exist to estimate the dispersion diagram of a waveguide from a measurement on the finite length waveguide (see [11] for example), we use a synthetic method [35] based on the measurement of the resonance frequencies of the cavity formed in Fig. 1. Such synthetic method utilizes a fitting algorithm to estimate the corresponding Bloch phase shift $k(\omega)d$ varying as a function of frequency.

This approach is also utilized in [36], for instance, to retrieve the dispersion of a slow-wave structure. As such, the dispersion relation takes the general form $D(\omega,k) = \sum_0^\infty a_m \cos[mk(\omega)d]$ with $a_m$'s coefficients are geometry-dependent and determined using the resonance frequencies of the structure (see equations (2)-(9) in [35]), and here the Bloch wavenumber is real-valued within the range $0 < k(\omega)d < 2\pi$. The dispersion is then approximated by taking into account the number of observed resonances in the finite-length structure, and the number of resonance is denoted by $M$. Thus the dispersion relation is approximately equivalent to [35]

$$D(\omega,k) \cong \sum_{m=0}^{M-1} a_m \cos[mk(\omega)d] \quad (1)$$

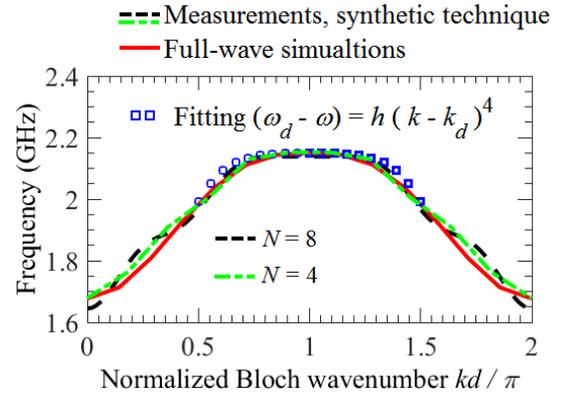

Fig. 5. Synthesized dispersion relation (1) of the waveguide in Figs. 1 and 3 based on reflection measurements of $S_{11}$, showing the propagation wavenumber versus frequency. This is in good agreement with the one obtained via full-wave simulations and with the characterizing DBE fitting dispersion law $(\omega_d - \omega) = h(k - k_d)^4$.

Considering the case of the DBE waveguide with $N = 8$, we have observed 9 resonances. Accordingly, we take $M = N+1$ in the above approximate dispersion relation (1). The resulting unknown coefficients $a_m$ are found to be $a_0 = 1.94$, $a_1 = -0.22$, $a_2 = -0.034$, $a_3 = -0.043$, $a_4 = 0.024$, $a_5 = -0.017$, $a_6 = 0.025$, $a_7 = -0.042$, and $a_8 = 0.032$. We report in Fig. 5 the comparison between full-wave simulation of the periodic waveguide using the eigenmode solver in CST and the corresponding measurement using the synthetic technique of the $N = 8$ cell cavity described above. A very good agreement between simulations and measurements is demonstrated in Fig. 5. We also superimpose the typical DBE dispersion relation $(\omega_d - \omega) = h(k - k_d)^4$ where $\omega_d = 2\pi(2.16\,\text{GHz})$, $k_d = \pi/d$, and $h$ is fitting constant. The fitting results in $h=435.57$ rad m$^4$/s, with an R-squared measure (refer to p.245 in [37] for the definition of the R-squared parameter) of 0.9994 indicating an excellent fit in the range $0.92 < f/f_d < 1$. It is worth mentioning that the synthetic technique assumes that single propagating mode (both forward and backward harmonics) is excited in the frequency range of interest [35], therefore the response is in very good agreement near the DBE in which only the overlapping (i.e., degenerate) DBE modes are present. On the other hand, some discrepancy between simulations and measurements away from the DBE is observed as seen in Fig. 5. The discrepancy is attributed to the possible excitation of another propagating mode (lower order mode exhibiting an RBE, that is propagating in the frequency range 1.6 GHz to 1.95 GHz as shown in Fig. 1(b)). We also perform the dispersion synthesis for the waveguide with $N = 4$ unit cells whose resonances frequencies are provided in Table I, and the dispersion shows very good agreement with the full-wave simulations as well the case with the measurement with $N = 8$ unit cells.

To provide yet another substantial evidence that the designed waveguide supports DBE, we report the very unusual scaling of the loaded quality factor $Q$ of the DBE resonance with the resonator's length. Here the estimated loaded quality factor is defined as $Q = 1/\text{FBW}$, where FBW is the fractional power bandwidth defined as the ratio between the 3-dB bandwidth (linewidth) of reflectance to the resonance frequency $f_{r,d}$ closest to the DBE (i.e., the closest to 2.16 GHz). We retrieve the FBW from measuring $|S_{11}|$ spectrum for different lengths of the waveguide at the resonance closest to the DBE, with one port excited while the other port is shorted. While making sure that the DBE resonance is critically coupled. The plot in Fig. 6 shows that the $Q$ factor grows as $N^5$ for large number of cells $N$, which is a clear indication of the presence of DBE occurrence as explained in [3], [10], [28]. It is important however to state that based on previous theoretical studies it is expected that for even longer waveguides, with $N \gg 8$, the $Q$ factor of the resonance near the DBE ceases to exhibit the $N^5$ growth trend owing to the losses of the waveguide copper walls and other tolerances ($Q$-factors for $N > 8$ are not shown in Fig. 6, but will be revisited in the next subsection). The loaded $Q$ factor measured from the FBW for the $N = 8$ unit cell waveguide is 1250 which agrees very well with the full-wave simulation results (simulated $Q \sim 1297$) as reported in Fig. 6.

It is also worth mentioning that the $Q$ factor, or equivalently the FBW, does not change significantly when changing the port 2 termination of the waveguide between open, short, and the 50 Ohm load conditions. To quantify that, the loaded $Q$ for the three loading cases depicted in Fig. 4(c) is 1250, 1283, and 1215 for the short, open and 50 terminations respectively, for $N=8$. This is also a direct consequence of the giant DBE resonance and $Q$-factor stability against loading discussed in [24], which is inferred from Fig. 4(c). Furthermore, as discussed in [19], [38], the $Q$ factor grows linearly in the log($Q$)-log($N$) plot in Fig. 6 even for small values of $N$ when the DBE feature is not yet evident. Indeed, for small values of $N$ the $Q$ factor can grow as $N$ or $N^3$, similar to a uniform cavity or a cavity with an RBE respectively, depending on the resonator terminations. On the contrary for larger $N$, i.e., $N > 4$ in our case, the $Q$ factor demonstrates a clear trend as $N^5$. However, in order to observe such high $Q$ feature of the resonance near the DBE for long waveguides, careful attention must be paid to the fabrication process of the waveguide in order to alleviate losses and maintain small tolerances that are detrimental to the DBE emergence. Importantly, in this paper we show that DBE can exist in a waveguide loaded with rings without an extremely sophisticated fabrication process. We stress that tolerances must be taken in to account when prototyping a specific application based on DBE in principle, which can be prone and sensitive to perturbations.

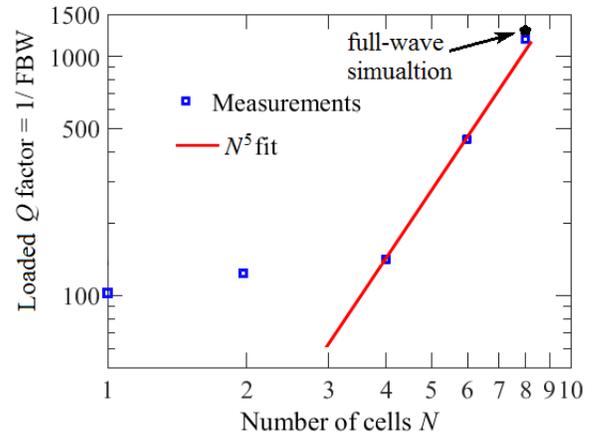

Fig. 6. The estimated loaded quality factor based on $Q = 1/\text{FBW}$ of the DBE cavity versus the number of unit cells $N$ retrieved from measurement of the $S_{11}$ response and FBW of the resonance at $f_{r,d}$. The scaling of the quality factor is fitted to $N^5$ for larger $N$ that is a feature of the DBE resonance. We also show the calculation for $N = 8$ using full-wave simulations reported in Fig. 3.

### C. Transmission measurements

We also report the transmission characteristics of the finite length waveguide by measuring the two port S-parameters of the waveguide by connecting the coaxial probes at both ports to a network analyzer, as seen in Fig. 7. There we report the measured reflection $|S_{11}|$ and transmission $|S_{21}|$ parameters near the DBE as well as the same values using full wave simulations. Good agreement is reported between simulations and measurements in the resonance frequency; however, the measured $|S_{21}|$ is ~ 4 dB lower than the simulated one and this is attributed to irregularities in waveguide walls, and losses in the external connectors and soldering spots not accounted for in the full-wave simulations. Note that the full-wave simulation shows a slight shift in the resonance frequency (~1 MHz) compared to the measurements. Better agreement of the resonance frequency in full-wave simulations and measurements can be achieved by adjusting the coaxial probe lengths and the rings misalignment angle. Indeed, when the probe length in simulation is adjusted to 23.5 mm while rings misalignment is set to ~58.5 degrees, full-wave simulation provides an exact fit of the resonance peak obtained from measurements (not shown in Fig. 7). In the present case the most remarkable features of DBE are not affected by such detuning of the misalignment angle and probe length. However, in general it is necessary to investigate the effect of such detuning and tolerances when designing structures that supports a DBE.

We also report in Fig. 8 the measured transmission group delay of the waveguide at the DBE resonance. The transmission group delay is defined as $\tau = -\partial(\angle S_{21})/\partial\omega$ where $\angle S_{21}$ is the phase of the S$_{21}$ scattering parameter and it is obtained by performing a numerical derivative of measured phase data. In Fig. 8 the transmission group delay is shown for different lengths of the waveguide, from measurements and full-wave simulations, normalized by the free space delay $Nd/c$. Measurements and simulations are in good agreement, and correlate to the scaling trend $N^4$ for large $N$ as predicted from theory [10]. We also show the measured group delay after dismantling and reassembling the $N = 8$ waveguide, shown by the star symbol in Fig. 8, which indicate that our results are easily reproducible in the same setup.

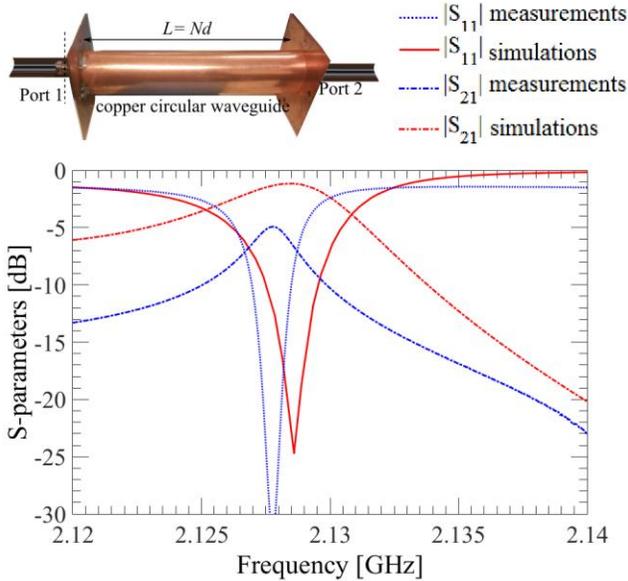

Fig. 7. Magnitude of the scattering parameters |S$_{11}$| and |S$_{21}$| obtained from both measurements and full-wave simulations versus frequency. The waveguide has $N = 8$ unit cells, and it is terminated on coaxial probes as shown in Figs. 1 and 4.

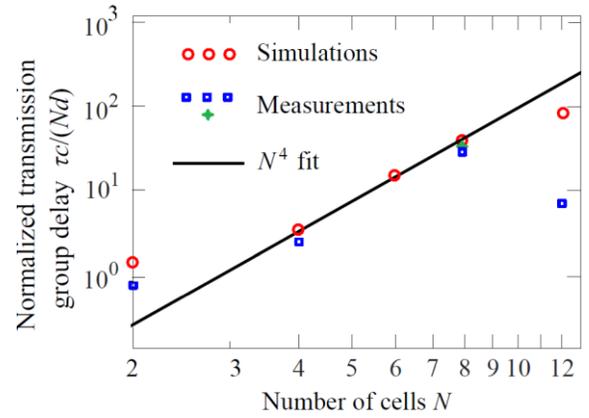

Fig. 8. Scaling of the normalized transmission group delay defined as $\tau = -\partial(\angle S_{21})/\partial\omega$ where $\angle S_{21}$ is the phase of the S$_{21}$ parameter, measured for different lengths of the waveguide. The result from full-wave simulation is also shown demonstrating good agreement with the measurements and the theoretical scaling trend $N^4$ for large $N$. However, measurements of group delay for $N=12$ are affected by losses that tend to limit the characteristic DBE features for long waveguides. The star symbols represent a different set of measurements after dismantling and reassembling the waveguide for $N = 8$.

Figure 8 also demonstrations that the group delay declines for large $N$ as losses and/or disorders dominate when the waveguide length increases as discussed earlier in the context of Fig. 6. In fact, the DBE transmission peak for $N = 12$ was completely deteriorated in the measured response corresponding to |S$_{21}$| < −40 dB at the DBE resonance of the finite length structure $f_{d,r} = 2.158$ GHz GHz when $N=12$. Note that this frequency pertains to a resonance at a frequency lower than that of the measured DBE $f_d = \omega_d/(2\pi) \approx 2.16$ GHz, therefore the strong attenuation is not due to the band gap that indeed occurs at slightly higher frequency, but due to losses and misalignment disorders. Interestingly, full-wave simulations also predict such decline of the loaded $Q$ factor as well as the group delay for the DBE resonance for long structures. Yet the group delay obtained from simulations is at least an order of magnitude larger than the one obtained from measurements. In other words, full-wave simulations either underestimate the effect of losses in the fabricated waveguide walls or the extra loss in the measurement is due to alignment tolerances, especially in long waveguide lengths. Note that the transmission coefficient is −40dB; since transmission at the DBE resonance in long waveguides is sensitive to losses, and the estimation of group delays may not be very accurate. However, the results clearly indicate that the band edge resonance's quality factor deteriorates more significantly than the lower order resonance proving that the DBE condition is extremely sensitive to losses and to external perturbation. Accordingly, a thorough analysis of perturbations such as detuning the misalignment angle, or altering the inter-ring spacing must be carried out in order to gain a deeper understanding of perturbations of such fourth order degeneracy associated with the DBE. Moreover, it is clear also that in DBE structures with losses an optimum number of unit cells shall be used such that the DBE feature is advantageously utilized. Incidentally, such perturbations can be used in tuning the response in active devices [17], [18] as well as in pulse compression schemes [24], which opens new frontiers for tunable applications such as antennas and reflectors, as well as such as pulse compressors or sensors.

DBE can be utilized in many applications since it provides for high quality factors resonators that can be easily detuned. DBE was also used in gain devices to obtain low threshold oscillators even when losses are present [18]; for generating high power at microwaves using an electron beam as discussed in [17], [18] or in a circuit with solid-state active components [25], [39]. As a final note, we again stress that the geometry in Figs. 1 and 3 is not the only possible waveguide implementation of DBE; this fourth order degeneracy can be found in many other loaded waveguides and using advanced fabrication technology and materials, for high power applications for instance [14], [15], in which the absolute values of the $Q$ factor can be even higher than the ones reported here.

## IV. CONCLUSION

We have reported for the first time an experimental demonstration of a fourth order degeneracy condition called degenerate band edge (DBE) in fully metallic circularly loaded waveguides in the S-band. Measurements of the retrieved dispersion relation, quality factor as well as the group delay were found to agree very well with full-wave simulations and follow the giant scaling features pertaining only to DBE characteristics. Such waveguide of finite length forms a resonator with very distinct characteristics. For example, we have shown that the resonance frequency and quality factor do not significantly change for large variations of the load. We also have shown that the loaded $Q$ factor for three termination conditions at the DBE transmission resonance in the order of 1207, 1250, and 1283 when port 2 is terminated with a "50-Ohm", "short" or "open", respectively. The DBE occurrence has been proved for the structure shown in this paper however DBE can be engineered in many other waveguide configurations and used to conceive devices for many applications such as pulse compression [24], oscillators [25], high power microwave source [6], [17], [18], filters, sensors and radio frequency high power modulators.

ACKNOWLEDGEMENT

The authors acknowledge useful discussions with Prof. A. Figotin, University of California, Irvine. Furthermore, they would like to thank CST of America, Inc. for providing CST that was instrumental in this work.